\documentclass[a4paper]{article}
\usepackage{graphicx}
\usepackage{INTERSPEECH2021}
\usepackage{subfigure}
\usepackage{multirow}
\usepackage[usestackEOL]{stackengine}
\usepackage{color,soul}
\usepackage[export]{adjustbox}
\usepackage{lineno,hyperref}
\title{Protecting gender and identity with disentangled speech representations}
\name{Dimitrios Stoidis, Andrea Cavallaro}

\address{
  Centre for Intelligent Sensing, Queen Mary University of London}
\email{dimitrios.stoidis@qmul.ac.uk, a.cavallaro@qmul.ac.uk}

\begin{document}

\maketitle
\begin{abstract}

Besides its linguistic content, our speech is rich in biometric information that can be inferred by classifiers.
Learning privacy-preserving representations for speech signals enables downstream tasks without sharing unnecessary, private information about an individual. In this paper, we show that protecting gender information in speech is more effective than modelling speaker-identity information only when generating a non-sensitive representation of speech. Our method relies on reconstructing speech by decoding linguistic content along with gender information using a variational autoencoder. Specifically, we exploit disentangled representation learning to encode information about different attributes into separate subspaces that can be factorised independently.
We present a novel way to encode gender information and disentangle two sensitive biometric identifiers, namely gender and identity, in a privacy-protecting setting.
Experiments on the LibriSpeech dataset show that gender recognition and speaker verification can be reduced to a random guess, protecting against classification-based attacks. 
\end{abstract}

\noindent\textbf{Index Terms}: privacy, soft biometrics, disentangled representation learning, variational autoencoder

\section{Introduction}\label{intro} 

The human voice provides a uniquely identifiable pattern for each individual, known as the voiceprint, a set of spectral characteristics that can be inferred and may be used for user profiling~\cite{userprofiling, singh}. 
Privacy-enhancing technologies aim to reduce the ability of an attacker to infer personal attributes when performing classification tasks, such as speech transcription~\cite{aloufi, vc1}. 

Approaches to protect voice privacy include voice transformation, voice conversion and synthesis, and representation learning. {Voice transformation}~\cite{voiceconvergindeid, hidebehind} modifies the non-linguistic information of speech to de-identitfy the speaker by applying basic transformations on features such as the pitch, time-scale components and energy modifications. 
Approaches that convert voices to {synthesised identities}~\cite{VC_adv} or produce new voices with generative networks~\cite{ericsson2020adversarial} learn to obfuscate sensitive attributes by synthesising new acoustic features mapped onto the linguistic content of speech.
{Representation learning} encodes each factor of variation in the data into separate compressed representations of latent subspaces~\cite{RLreview}. These compact representations enable individual component manipulation for speech reconstruction with a certain level of privacy~\cite{aloufi, vc1, AdvRL}. Furthermore, adversarial representation learning can be used to hide the identity of a speaker~\cite{AdvRL}. 

In this paper, we propose a new way to encode gender by creating an embedding space that independently captures gender information, and show that considering gender information improves the privacy-utility trade-off compared to existing representations. We consider attacks targeted at extracting identity and gender, when speech recognition is the downstream task. Within this scenario, we hypothesise gender to be the primary component in the auditory identity of a speaker and that {neutralising} gender also affects the ability to recognise the identity.

To this end, we define a generative model that allows us to independently modify individual attributes in a lower dimensional space. To mitigate undesired attribute inferences, we model gender information in two settings: (i) by using gender information alone and (ii) by including both gender and speaker-identity information. Specifically, we investigate the extent to which concealing information related to a soft biometric identifier such as gender helps concealing identity information. To the best of our knowledge no prior work focused on disentangling identity and gender from content in order to use gender information as a proxy for guiding privacy-preserving voice representation learning. 
We show that our method is more effective in protecting the identity and gender of a speaker from inference than existing approaches on disentangled representation learning that use only identity information to perform the anonymisation.

\section{Related Work}\label{related_work}

Speech data anonymisation methods include voice transformation~\cite{hidebehind}, voice conversion and speech synthesis~\cite{VC_adv, ericsson2020adversarial, x-vecan}, and representation learning. Representation learning can in turn be classified as adversarial~\cite{AdvRL}, disentangled~\cite{aloufi, vc1, Wu} or hybrid~\cite{noe}.

{Voice transformation} aims at modifying the non-linguistic components of an utterance, leaving the textual content unaffected. Voice transformation involves extracting learned features, such as x-vectors~\cite{x-vecan, tomashenko2020introducing}, or hand-crafted features, such as fundamental frequency (F0), Mel-Frequency Cepstral Coefficient (MFCC) and energy. The transformation typically refers to modifying instantaneous characteristics of the source signal such as the pitch, spectral envelope, time-scale and energy~\cite{voiceconvergindeid}.
An audio sanitiser may transform voice characteristics by using frequency warping~\cite{hidebehind}.

{Voice conversion and speech synthesis} map the input voice of a source speaker into that of a target speaker~\cite{VC_adv, siamese_ae}.
These methods take advantage of the underlying statistical characteristics of sounds to discriminate between different aspects of audio signals and disentangle the textural style from the semantic content~\cite{style}.

{Representation learning} facilitates the task of extracting useful features from the data, usually with autoencoding architectures that compress the data to a smaller dimensionality~\cite{AdvRL, noe}.
Disentangled representations are meant to capture the different independent factors of variation in the data~\cite{aloufi, vc1, Wu}.

A few works consider gender as an attribute to protect.
PCMelGAN~\cite{ericsson2020adversarial} uses an adversarial approach by generating speech with generative adversarial networks (GANs)~\cite{gans}.
GANs, Variational Autoencoders (VAEs) and their combination can be used to protect gender information through voice conversion with a disentanglement approach~\cite{Wu}.
Noe et al.~\cite{noe} disentangle gender information from x-vectors with adversarial learning. Low, moderate and high privacy levels may be defined using a disentangled representation~\cite{aloufi}, where the moderate level aims to reduce the success rate of inferring sensitive attributes (emotion and gender) while maintaining high accuracy in inferring speaker identity and linguistic content.

Beyond privacy, VAEs have been used to 
capture independent factors of speech at different timescales in an unsupervised way~\cite{Kingma}. Such a {disentangled} representation is divided into discrete and independent subspaces, such that each subspace corresponds to only one factor of variation in the data, without affecting, or being affected by the transformation of the other subspaces~\cite{RLreview}.

The Factorised Hierarchical VAE (FHVAE)~\cite{hsu2017FHVAE} uses multi-scale approach to manipulate different sets of latent variables by factorisation and discriminates between sequence-level (speaker identity) and segment-level (content) attributes through a hierarchical process.
An auxiliary lower bound designed to disentangle global from local information at different time-scales is introduced in AuxVAE~\cite{AuxVAE}, showing that the latent space can capture individual speaker information.
Similarly, VQ-VAE~\cite{oordNDRL} produces an encoded representation of speaker-specific information using vector quantisation.

\section{Method}\label{method}

\subsection{Privacy scenario}

We consider the identity and the gender of a speaker as the private attributes to protect, and speech recognition as the target task. We view speaker identity in the context of a speaker verification task, which consists in determining whether an utterance corresponds to the utterance of an enrolled speaker. 

We assume that the attacker has anonymised speech samples with labels and attempts to infer the gender and identity of the speaker by classification and verification, respectively. We consider an \emph{ignorant} attacker with no knowledge of the target identity used for conversion or of the architectural design of the Encoder-Decoder network~\cite{vc1}.

We quantify {privacy} as the degree of randomness in classifying the gender and/or verifying the identity of the speaker. We measure {utility} by the speech recognition accuracy.

\subsection{Disentanglement approach}\label{approach}

We hypothesise that the linguistic and biometric information that is entangled in speech signals can be decomposed into three subspaces, namely {content} (the non-sensitive subspace), {identity} and {gender} (the sensitive subspaces). Thus we aim to learn disentangled representations of content, identity and gender, such that gender and identity attributes cannot be inferred given a set of reconstructed speech utterances of a speaker.

We use an encoder network consisting of a one-dimensional convolutional layer followed by 5 blocks of linear layers along with Vector Quantisation (VQ) to obtain a quantised latent space encoding linguistic content~\cite{oordNDRL}.
We adapt the disentangled representations method between content and speaker identity~\cite{aloufi} to disentangle the gender component.

We use Vector Quantised VAE (VQ-VAE)~\cite{oordNDRL} to extract the linguistic content by producing a discrete latent space through vector quantisation.
Using quantised instead of continuous latent vectors can facilitate the disentanglement of non-sensitive and sensitive speech components that can be represented with discrete sequences.

Vector quantisation in the latent space acts as a reparameterisation trick that discretises the encoder output into fixed size vectors.
The {encoder network} parameterises a posterior distribution $q(z \vert x)$, where $x$ is the input speech signal and $z\tt\sim Z$ the vector of latent variables sampled from space $\tt Z$ according to a uniform prior distribution $p(z)$.
In order to disentangle variables associated with content, identity and gender, we consider the latent variables $z$ representing the non-sensitive information (content) and the vector of latent variables $b \tt\sim B$ representing sensitive information to be protected with $b_j\in b$ and $j\in \{1,2\}$ representing gender and identity.
Then $z$ and $b$ are said to be disentangled if they can be factorised as~\cite{fairRL}:
\begin{equation}
    q(z,b)=q(z)\prod_j q(b_j),
\end{equation}
where $q(.)$ represents the posterior of each vector of variables and $q(.,.)$ represents their aggregate posterior.

Disentangled representations help limit the amount of shared information between separate latents and thus encourage predictions to depend only on relevant (non-sensitive) latent dimensions~\cite{locatello}.
Thus, we create an embedding space that uniquely defines gender and use it to reconstruct speech, either alone or along with identity. 
By independently encoding sensitive and non-sensitive information and combining each of them by concatenation in the decoding stage, we manage to produce disentangled representations of speech.

\begin{figure}
    \centering
    \includegraphics[scale=0.2]{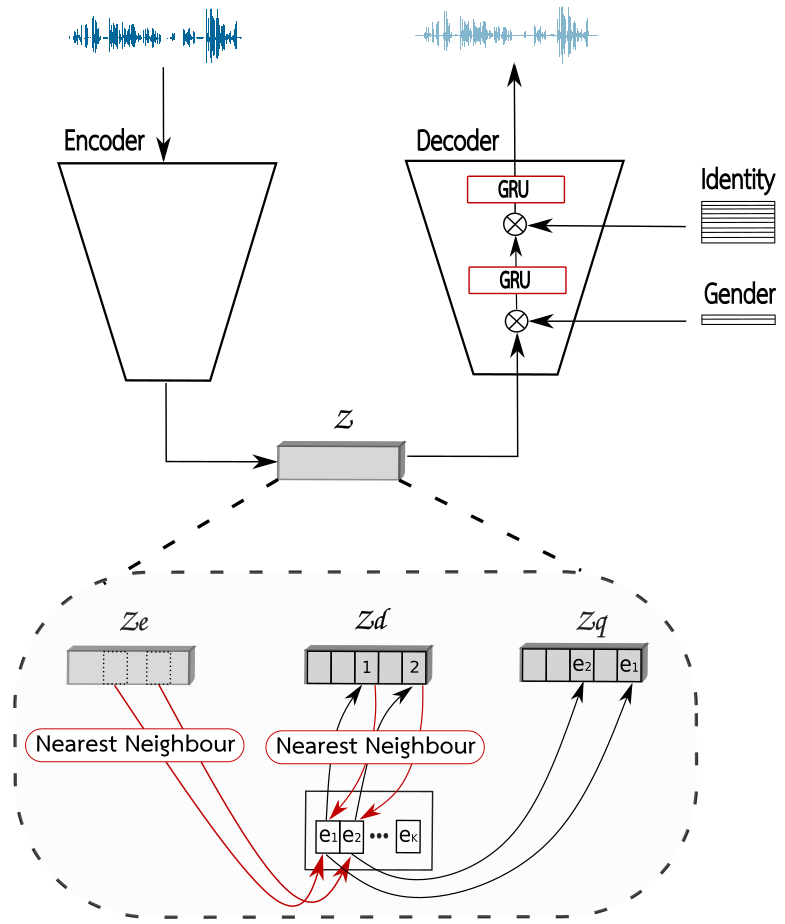}
    \caption{Our architecture to encode content and reconstruct speech with the desired speaker identity and/or gender. 
    The latent space $z$ is reparameterised to encode content information.
    The output of the encoder $z_e$ is first discretised into $z_d$ as a one-hot vector of indices by nearest neighbour and then mapped onto each embedding vector $e_i$ by nearest neighbour look-up to give the quantised latent content vector $z_q$ fed as input to the decoder for reconstruction. The decoder (vocoder) generates speech by mapping the speaker identity and gender onto the content or by selectively mapping each attribute separately. The identity and gender embeddings are concatenated with the quantised content information through two gated recurrent units (GRU).}
    \label{fig:network}
\end{figure}

The vector quantisation step involves learning a latent embedding space (or codebook) $e \in \mathcal{R}^{K\times D}$, where $K$ is the number of embedding vectors (or codewords) and $D$ is the length of each vector.
The vector of latent variables $z_e$ taken as output from the encoder is discretised into $z_d$ by nearest neighbour look-up to the closest codeword $e_i$, where $i \in 1, 2, ..,K$ (see Fig.~\ref{fig:network}).
This models the posterior distribution $q(z \vert x)$ as a one-hot vector of indices, where the closest entries are assigned an index $k$ corresponding to the codeword $e_k$, and 0 at all other positions, essentially applying a discretisation bottleneck.

We obtain the input to the decoder, $z_q$, by mapping the nearest corresponding embedding vector $e_K$ onto the one-hot vector.
In this way, the content is encoded onto a discrete latent space, defined by a codebook consisting of $K$ codewords.
To reconstruct audio waveforms from the acoustic features, instead of using a traditional vocoder such as Griffin-Lim~\cite{griffin1984signal},
we use -- for its improved quality of the generated speech -- a modified version of WaveRNN~\cite{wavernn}, where two gated recurrent units (GRU) are used to capture content information~\cite{oordCPC}.
The logits that go through the softmax layer to predict an 8-bit $\mu$-law sample are taken as output of two linear layers following the autoregressive GRU cells. 

To condition the decoder to reconstruct the speech signal with gender characteristics, we use a single stacked bi-directional GRU that takes as input the concatenation of the discrete latent vector output $z_q$ along with the gender embeddings.
We modify the dimensionality of the embedding space used to model speaker information~\cite{oordCPC} and adapt it to encode gender information.  
While, speaker information is encoded by 64-dimensional vectors corresponding to each speaker in the set, only two of these vectors are needed in a binary gender encoding, reducing the dimensionality needed.
To combine speaker identity and gender information, we add a bi-directional GRU layer to model speaker information.
The input to this recurrent layer is taken as the output of the first conditional GRU, concatenated with the speaker embeddings.
We increase the input channels by four times to account for the dimensionality produced by the concatenation of content and gender passed through a bi-directional GRU.

The architecture defined above allows us to consider five settings: Same Identity (SI); Random Identity (RI); Random Gender (RG); Same Identity, Random gender (SIRG); and Random Identity, Same Gender (RISG). With SI, speech is reconstructed maintaining the identity of the source speaker. The goal is to reduce gender inference while preserving speaker identity. With RI, speech is reconstructed using the encoded identity of another speaker in the set, chosen at random. With RG, speech is generated by mapping a gender at random onto the chosen utterance. With SIRG, speech is generated using the same identity as the source speaker while varying the mapped gender at random for each utterance. Thus, the utterances of a speaker will be mapped first onto the same identity and then onto a randomly chosen gender. Finally, with RISG, speech is generated by mapping a randomly chosen identity in the set, followed by mapping to the same gender as the source speaker. 

Note that the last three settings consider gender information and are specific to our scenario, whereas existing works consider only identity information. In the next section, we evaluate these five settings and their impact on privacy and utility.

\section{Validation}
\label{results}

\subsection{Experimental setup}

We compare the proposed approach with methods that use disentangled representations in a privacy setting and consider speaker identity and/or gender as sensitive attributes. The methods we consider are EDGY~\cite{aloufi}, Disentangl.-based VC~\cite{vc1}, and the VAE model used by Wu et al.~\cite{Wu}, which we will refer to as Client-VAE, that uses identity information during training.
We use the publicly available code for EDGY~\cite{aloufi}, and the results reported in the corresponding papers for Disentangl.-based VC~\cite{vc1} and Client-VAE~\cite{Wu}.

We conduct three sets of experiments, one that uses only gender information, and two that include gender and identity information.
To facilitate the comparison, we choose methods whose results were presented on the same dataset, LibriSpeech~\cite{librispeech}, which consists of read speech from audiobooks. 
We use the 100-hour training set consisting of clean speech and 251 speakers as it has balanced gender distribution.

\subsection{Performance measures}

We measure utility as word error rate (WER), which counts the number of insertions, deletions and substitutions between reference and predicted transcript, over the total number of words in the reference transcript. 

We measure the effectiveness of gender inference as the standard binary classification accuracy, which is the sum of true positives and true negatives over the total number of predictions. 

Finally, for speaker verification we consider as rate of failure of an attacker the equal error rate (EER). 
For biometrics verification (which favours a false rejection over false acceptance)
the lower the EER, the higher the accuracy of the speaker verification task. Instead, for a privacy scenario, an EER approaching 50\% is desired as it represents randomness in verifying the speaker identity~\cite{vc1,tomashenko2020introducing}.

\subsection{Models}

For {speech recognition}, we use DeepSpeech2~\cite{deepspeech}, which consists of 11 layers of alternating convolutional and recurent layers.
The model uses the Connectionist Temporal Classification (CTC) loss~\cite{graves2006ctc} 
to predict the most likely transcription and reported a word error rate (WER) of 6.75\% using a 3-gram language model on the LibriSpeech clean test set. 

For {gender classification}, we design a deep convolutional network with 5 stacked one-dimensional convolutional layers followed by batch normalisation and max polling with ReLU non-linear activation.
The output layer consists of a fully connected layer that outputs a pair of predictions for each binary class, passing through a sigmoid function.
We pre-trained the gender classifier network on the LibriSpeech train set for 50000 iterations with a learning rate of 0.0001 and the Adam optimiser, updating the weights using the cross-entropy loss function. We tested the gender classifier on the LibriSpeech clean test set and reported an accuracy of 91.37\%.

For {speaker verification}, we extract speaker-identifying features with Thin ResNet-34~\cite{resnet}, a smaller version of the original ResNet-34. The model was pre-trained on the VoxCeleb2~\cite{voxceleb2} dataset using the angular prototypical loss~\cite{resnet} and reported an EER of 5.73\% in our experiments when tested on the LibriSpeech clean test set.
We use a pre-trained encoder with Contrastive Predictive Coding (CPC)~\cite{oordCPC} with a codebook size of $K=512$ and $D=64$ on the ZeroSpeech 2019~\cite{zerospeech} dataset.

\subsection{Discussion}

\begin{figure}[t]
    \includegraphics[height=4.5cm, width=10cm, center]{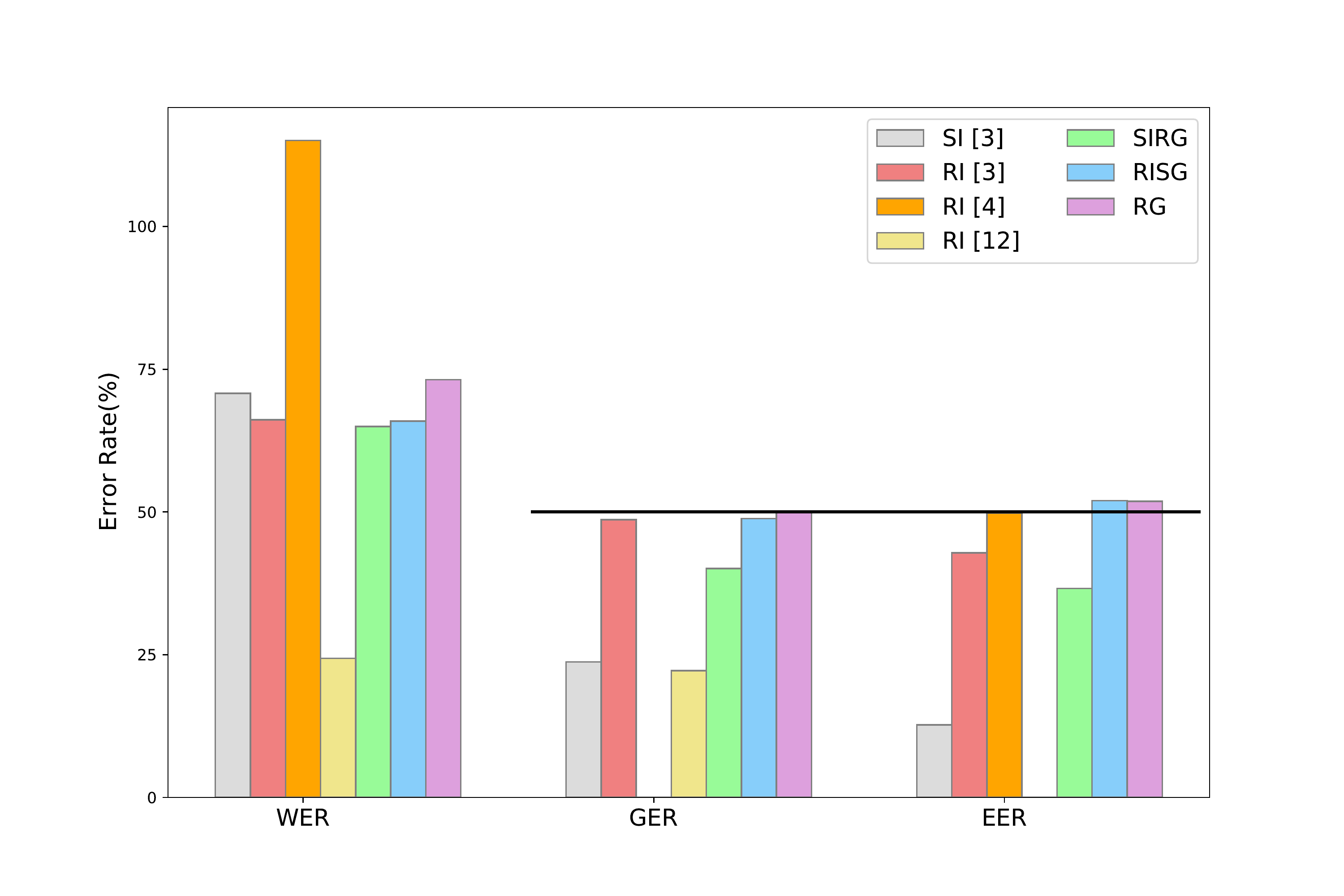}
    \caption{Visualisation of the utility and privacy levels obtained by different disentangled representations. Key -- SI: Same Identity; RI: Random Identity; RG: Random Gender; SIRG: Same Identity, Random Gender; RISG: Random Identity, Same Gender; WER: word error rate, GER: gender error rate; EER: equal error rate. Note that: GER is computed as complementary of the classification accuracy, i.e. ($100-Acc$); and GER and EER values close to 50\% (shown by the horizontal line) denote better privacy.}
    \label{fig:results}
\end{figure}

Table~\ref{tab:results} compares the recognition accuracy results for the methods under analysis with the settings defined in Sec.~\ref{approach}. Furthermore,  to visually  compare error rates for utility and privacy and to facilitate the identification of the appropriate trade-off,
Fig.~\ref{fig:results} shows a bar plot with the data of Table~\ref{tab:results}, but considering the Gender Error Rate, computed as $100-Acc$ where $Acc$ is the classification accuracy.

The SI setting provides an intermediate privacy protection as it is aimed at maintaining the speaker identity and ignores gender information (note that this setting corresponds to the \emph{moderate} privacy setting in EDGY). This setting shows that mapping onto the same identity induces distortion in the recognised identity, but remains quite low at 12.7\% EER, while the gender attribute can be inferred with high accuracy (70.78\%).

The goal of RI is to protect from inference attacks both identity and gender attributes in EDGY, while only identity is considered in Disentangl.-based VC and Client-VAE.
In the case of Disentangl.-based VC we consider the \textit{ignorant} scenario, where the attacker has no knowledge of the type of conversion being applied.
In RI, the speaker identity is better protected with Disentangl.-based VC than with other methods, at the cost of a higher utility degradation. 
In this setting, gender protection is also achieved in EDGY when converting to a random target identity, with a minimal discrepancy from a random guess (1.36 percentage points). 
A higher utility is achieved by Client-VAE, which however fails to protect from gender inference (classification accuracy of 77.8\%).

The settings that are specific to our methods (also) consider gender information.
In SIRG, adding random gender information with respect to SI affects both speaker and gender recognition rates, which are closer to random guesses.
Thus gender information is confirmed to play an important role for the auditory identity of an individual.
However, this setting is not as effective as using a random speaker identity, as shown for RI and RISG.
Although gender recognition remains the same in RI and RISG, speaker verification accuracy is reduced when adding gender information.
SIRG and RISG show that the encoded identity information is dominant over gender, due to increased dimensionality of the embeddings, as seen by the low EER in SIRG. 

Finally, RG offers better privacy protection both in terms of speaker verification and gender classification, compared with methods encoding identity information only.
With RG, both gender binary accuracy and EER are very close to the value of the ideal random guess.
Furthermore, our settings RISG and RG provide the best trade-off when both utility and privacy objectives are considered.
Also, the settings using speaker identity (SI and SIRG) help quantify the extent to which gender information is present when encoding identity information.
However, this does not hold true when encoding gender information alone, suggesting that gender can be disentangled from identity. Thus, using gender information is preferable when defending against inference attacks aiming at inferring the identity and the gender of a speaker.

\begin{table}[t]
    \centering
        \caption{Comparison of utility and privacy levels obtained by different disentangled representations. Key -- WER: Word Error Rate (\%); Acc: binary classification accuracy (\%); EER: equal error rate (\%); SI: Same Identity; RI: Random Identity; RG: Random Gender; SIRG: Same Identity, Random Gender; RISG: Random Identity, Same Gender. Note that: smaller values denote higher utility for WER and lower privacy for EER; larger values denote lower privacy for gender accuracy; and a value of WER $>$ 100 denotes insertion of new words in the predicted transcription. 
        }
    \begin{tabular}{lcccc}
    \hline
     {\bf Ref.} & \multicolumn{1}{c}{\bf Settings} & {\bf Utility} & \multicolumn{2}{c}{\bf Privacy} \\
       & & WER & Acc & EER \\
    \hline
    EDGY~\cite{aloufi}  & SI     & 70.78 & 76.27      & 12.70\\
    EDGY~\cite{aloufi}  & RI     & 66.17 & 51.36      & 42.85\\
    Disentangl.VC~\cite{vc1}   & RI     & 115.10& \textendash & 49.85\\ 
    Client-VAE~\cite{Wu} & RI  & 24.38 & 77.80 & \textendash\\ \hline
                    & SIRG &  64.99 &  59.89 & 36.63 \\
          ours      & RISG &  65.92 & 51.15 & 52.00\\
                    & RG   & 73.16 & 50.01      & 51.88\\
    \hline
    \end{tabular}
    \label{tab:results}
\end{table}

\section{Conclusions}

In this paper we showed that learning disentangled representations of speech for content, gender and identity enables the protection of user privacy from attribute inference attacks.
We also showed that, in order to protect the identity of a speaker, it is preferable to map the content onto a randomly chosen gender rather than a randomly chosen identity. 
By encoding gender information, we can better protect speaker information from inference using a state-of-the-art speaker verification model.
Specifically, mapping gender information onto the encoded content provides better privacy than using speaker information alone when reconstructing speech signals, thus improving upon the current state-of-the-art models for disentangled representations.

As VQ-VAEs are effective in disentangling latent spaces but lack in reconstruction capabilities compared to continuous VAEs, future work includes investigating representations that improve control over the degradation of the WER.
Furthermore, we will extend the current model and test its robustness when the knowledge of the attacker increases to cover the type of conversion used to protect the privacy of the speaker.\\*[0.5cm]

\noindent\textbf{Acknowledgements} We thank the Alan Turing Institute (EP/N510129/1), which is funded by the U.K. Engineering and Physical Sciences Research Council, for its support through the project PRIMULA.

\bibliographystyle{IEEEtran}

\bibliography{template}

\end{document}